\pdfoutput=1

\documentclass[twocolumn,twocolappendix]{aastex631}
\usepackage{here}
\usepackage{url}
\usepackage{color,ulem}

\defcitealias{2024ApJ...961...32K}{Paper I}
\defcitealias{2021ApJ...906L...3T}{T+21}
\shorttitle{Circumstellar structure of Tycho's SNR--broadband emission}
\shortauthors{Kobashi et al.}
\graphicspath{{./}{figures/}}

\begin{document}

\title{Exploring the circumstellar environment of Tycho's supernova remnant. II. Impact on the broadband non-thermal emission}

\author[0009-0008-4215-1049]{Ryosuke Kobashi}
\affiliation{Department of Astronomy, Kyoto University, Kitashirakawa, Oiwake-cho, Sakyo-ku, Kyoto 606-8502, Japan}

\author[0000-0002-2899-4241]{Shiu-Hang Lee}
\affiliation{Department of Astronomy, Kyoto University, Kitashirakawa, Oiwake-cho, Sakyo-ku, Kyoto 606-8502, Japan}
\affiliation{Kavli Institute for the Physics and Mathematics of the Universe (WPI), The University of Tokyo, Kashiwa 277-8583, Japan}

\author[0000-0002-4383-0368]{Takaaki Tanaka}
\affiliation{Department of Physics, Konan University, 8-9-1 Okamoto, Higashinada, Kobe, Hyogo 658-8501, Japan}

\author[0000-0003-2611-7269]{Keiichi Maeda}
\affiliation{Department of Astronomy, Kyoto University, Kitashirakawa, Oiwake-cho, Sakyo-ku, Kyoto 606-8502, Japan}

\correspondingauthor{Ryosuke Kobashi}
\email{kobashi@kusastro.kyoto-u.ac.jp}



\begin{abstract}
While the environment around Tycho’s supernova remnant (SNR) has long been believed to be close to homogeneous, the latest analysis of \textit{Chandra} data has identified a substantial deceleration of the forward shock which poses a major challenges to this picture. \citet{2024ApJ...961...32K} showed that the existence of dense molecular cloud (MC) surrounding a rarefied wind-like circumstellar matter (CSM) can explain this observational finding in term of the shock-expansion dynamics, supporting the so-called single-degenerate scenario for the progenitor system. 
We here extend this work to study the non-thermal emission processes and investigate whether such an environment is consistent with the observed multi-wavelength spectrum. While the simulated broadband spectrum based on the wind-MC environment is largely consistent with observations, we find that such an environment predicts a harder gamma-ray spectrum than observed due to the relatively low CSM density in the cavity interior of the MC. This difference can be at least partially attributed to the present one-dimensional setup of the model which does not account for the clumpy and multi-dimensional structure of the CSM. Our model provides predictions for the long-term evolution of the broadband spectrum which can be used to further probe Tycho's surrounding environment in the future, a key to resolving the long-standing issue of type Ia supernova progenitor channels.
\end{abstract}



\section{Introduction} \label{sec:intro}
Tycho's supernova remnant (Tycho's SNR, SN1572, G120.1+1.4; hereafter Tycho) is widely believed to originate from a thermonuclear SN explosion in the year 1572. 
The SN explosion appears to be a normal type Ia in terms of properties such as the light echo \citep{2008Natur.456..617K}, thermal X-ray emission from the ejecta \citep{2006ApJ...645.1373B}, the almost spherical morphology \citep[e.g.,][]{1997ApJ...491..816R} and so on. 
From the morphology and observed multi-wavelength spectrum, most previous data analyses and simulations support 
a homogeneous ambient medium for Tycho \citep[e.g.,][]{2005ApJ...634..376W,2006ApJ...645.1373B,2012AA...538A..81M,Slane2014,2019ApJ...876...27Y}. 

On the other hand, recently a non-uniform cavity surrounded by dense clouds has been suggested for Tycho's environment by multiple studies, such as by X-ray observations using \textit{Chandra} \citep{2016ApJ...823L..32W}, and possible association with dense clouds from infrared observation with \textit{Spitzer} \citep{2013ApJ...770..129W} and CO observation with the \textit{IRAM} 30~m telescope \citep{2016ApJ...826...34Z}. 
%
More recent proper motion measurements of the shock-expansion dynamics have further supported the existence of such a non-uniform cavity-like environment. In particular, a substantial deceleration of Tycho's forward shock (FS) is discovered by \citet{2021ApJ...906L...3T} with \textit{Chandra} (hereafter \citetalias{2021ApJ...906L...3T}), from which a low-density CSM (cavity) surrounded by a dense molecular cloud (MC) has been inferred. 
Based on the proper motion data from \citetalias{2021ApJ...906L...3T}, \citet{2024ApJ...961...32K}\citepalias[hereafter][]{2024ApJ...961...32K} calculated the forward shock (FS) radius evolution using one-dimensional hydrodynamic simulations for various models for the environmental properties.
This work suggested that a wind-MC structure for Tycho's environment, i.e., (i) a wind-like ($\rho(r)\propto r^{-2}$) and anisotropic CSM with a relatively low density and (ii) a dense environment surrounding such a CSM, can satisfactorily explain the shock-expansion data (see also \citet{2013MNRAS.435.1659C} for a similar model). The existence of such an environment supports a single-degenerate scenario for Tycho. However, whether the CSM proposed in \citetalias{2024ApJ...961...32K} can explain the broadband electromagnetic spectra observed so far still remains to be addressed. This study targets on making a first quantitative investigation of this consistency between the latest picture for Tycho's environment and the non-thermal spectra collected from available multi-wavelength observations.  


In Section~\ref{sec:setup} we introduce the setup of our hydro and non-thermal emission models. Section~\ref{sec:res} illustrates our main simulation results and a comparison among different models will be provided. Section~\ref{sec:disc} explains the possible multi-dimensional effects which are not included in our model, and addresses the possible impacts of such effects on our results. A few additional caveats are also presented there. Section~\ref{sec:sum} summarizes our work.


\section{Setup and models}\label{sec:setup}
\begin{figure}[ht]
    \epsscale{1.15}
    \plotone{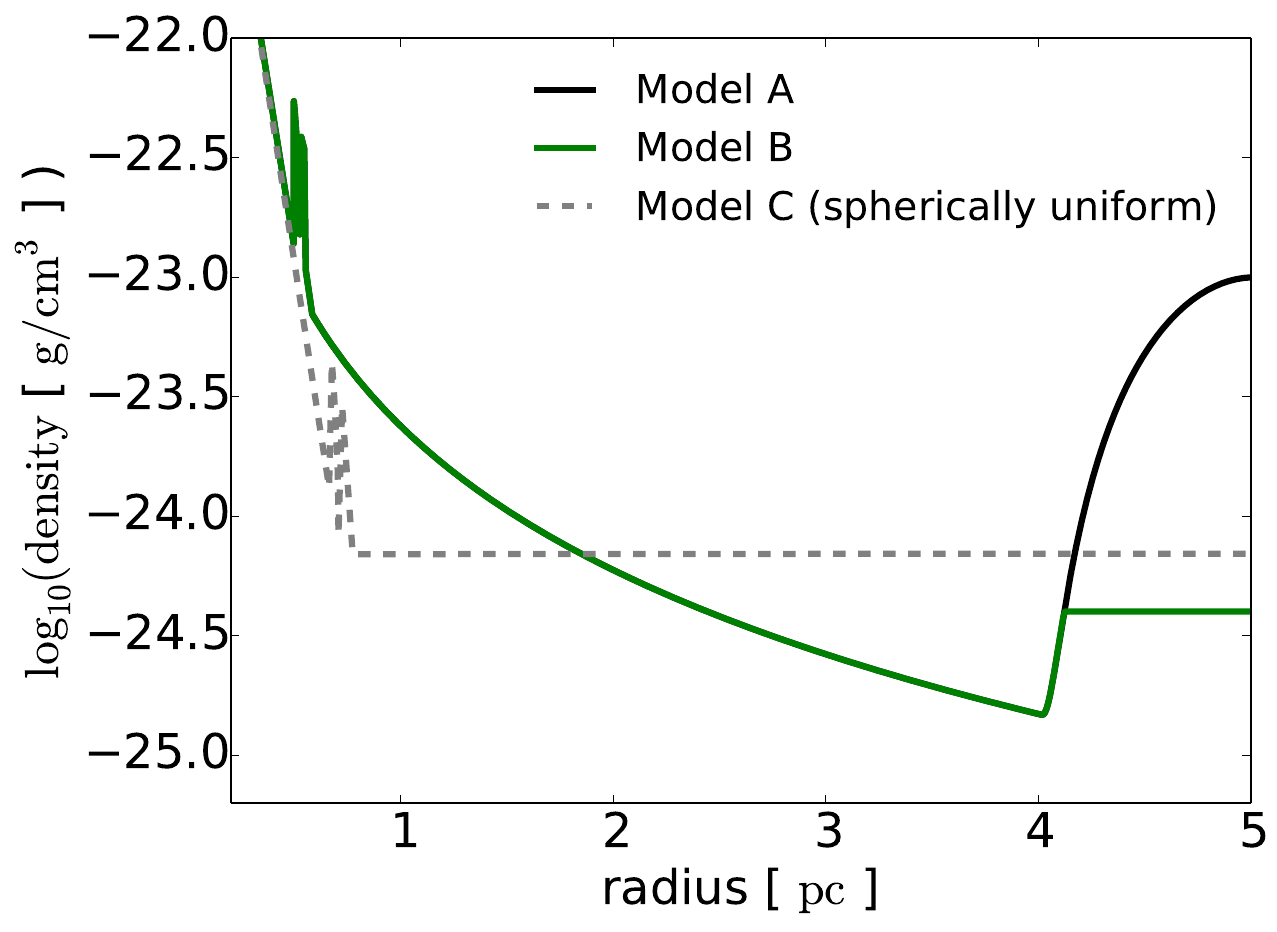}
    \caption{The initial density profile of Models A (black solid line; our fiducial model), B (green solid line; modified in a cloud region), and C (grey dashed line; spherical uniform model with the density $0.3 m_\mathrm{p}$ g cm$^{-3}$) for Region 13 at an age of 30 yr.}
    \label{fig:denden-init}
\end{figure}

We use the \textit{CR-Hydro} code developed by \citet{2019ApJ...876...27Y} and \citet{2022ApJ...936...26K} (see also references therein) to calculate the hydrodynamic evolution as well as the accompanied broadband non-thermal emission. The \textit{CR-Hydro} code is based on a one-dimensional hydro code \textit{VH-1} which solves the time evolution of an SNR on a Lagrangian grid \citep[e.g., ][]{2001ApJ...560..244B}, coupled to a semi-analytic calculation of nonlinear diffusive shock acceleration (NLDSA) of non-thermal particles. \textit{VH-1} has also been used previously for the investigation of Tycho's environment in \citetalias{2024ApJ...961...32K}. 
We divide Tycho's SNR into 13 azimuthal regions (Regions 1--13, numbered in an order from the north in the CCW direction as shown in Fig.~1 of \citetalias{2021ApJ...906L...3T} or Fig.~1 of \citetalias{2024ApJ...961...32K}) in the 2-D projection plane, and model the one-dimensional (radial) evolution separately in each azimuthal region with the \textit{CR-Hydro} code. We then calculate the associated age-specific non-thermal emission spectrum for each region and integrate them volumetrically to either compare with the observed spectral data or make predictions for future observations. 
Unless otherwise specified, we employ the geometric center \citep{2005ApJ...634..376W} for the explosion center (instead of the pressure center calculated from overall pressure gradient \citep{2016ApJ...823L..32W}) based on the recent study of 3-D ejecta properties \citep{2022ApJ...937..121M}. 
For the volumetric integration, we calculate the total emission fluxes by summing the flux $f_E(\psi)$ at a certain photon energy $E$ over the azimuth $\psi$,
\begin{equation}\label{eq:sum-up}
f_{E,tot}=\int_{-1}^{1}d(\cos(\psi-\delta\psi))\ \frac{f_E(\psi)}{4\pi}\cdot2\pi,
\end{equation}
where the offset azimuth is chosen to be $260.3^\circ$ (roughly East to West), which defines an apparent axis of symmetry in the morphology of the Tycho's SNR in the two-dentinal projection \citepalias{2024ApJ...961...32K}. Accordingly, the modeled spectra in the northern hemisphere are taken as being representative in integrating the spectra in the $\psi$-direction. We have also checked the $\psi$-integration with $\delta\psi=78.4^\circ$ (for thich the models in the southern hemisphere are used); we confirmed that it does not make significant difference (Fig.~\ref{fig:res-1sed}d). The term $2\pi$ comes from the assumed axial symmetry against the line $\psi=\delta\psi$. The incorporation of information from the velocity-field distribution along the line-of-sight \citep[e.g.,][]{2024ApJ...962..159U}\footnote{We note that \citet{2024ApJ...962..159U} has confirmed that the velocity distribution along the line-of-sight is qualitatively similar to what we found in the projected plane.} is postponed to a future followup study.

For each azimuthal direction, we define a `fiducial' model for the circumstellar environment (Model A) using the best-fit hydro models derived in \citetalias{2024ApJ...961...32K} to fit the data obtained by the proper motion study by \citetalias{2021ApJ...906L...3T}. As explained above, this model features a gas density structure composed of a low-density wind cavity surrounded by a MC-like dense exterior region. In Model A, we simply extrapolate the gas density profile beyond the radius of the shock  in the year 2015 ($R_{2015}$) due to a lack of observational constraint from the currently available proper motion measurements. To quantify the impact of such an assumption,
we prepare a modified version (Model B) in which $\rho(R>R_{2015})=\mathrm{const.}$ 
Another Model C features a setup with a spherically uniform ambient density of $\rho = 0.3 m_\mathrm{p}$ g cm$^{-3}$, which can be compared to the results from previous works, e.g., \citet[][]{Slane2014, 2019ApJ...876...27Y}. 
The density structures of these models are shown in Figure~\ref{fig:denden-init}.

The key parameters for the particle acceleration calculation 
include the followings.
$\chi_\mathrm{inj}$ is the injection parameter related to the injection efficiency $\eta\propto\chi_\mathrm{inj}^3 e^{-\chi_\mathrm{inj}}$ which originates from the so-called ``thermal leakage'' model \footnote{Our result however does not depend on the particular assumption on the injection model; in practice, the injection efficiency $\eta$ is treated essentially as a free parameter to be assessed.}; $K_\mathrm{ep}$ is the electron-to-proton number ratio at relativistic energies; $\sigma_\mathrm{w}\equiv\frac{(B^2/8\pi)}{(\rho V_\mathrm{w}^2/2)}\sim10^{-2}$ is the wind magnetization parameter; $f_\mathrm{alf}\in(0,1)$ parameterizes the spatial variation of the Alfv\'{e}nic speed where $f_\mathrm{alf}=0$ corresponds to the case of $v_A(x)=v_{A,0}$ 
\citep[for details, see e.g.,][]{LEN2012,2019ApJ...876...27Y,2022ApJ...936...26K}. 
%
The non-thermal emission components considered in the spectral models include inverse Compton (IC) scatterings, synchrotron radiation, non-thermal bremsstrahlung emission, and $\pi^0$-decay \citep[][and references therein]{2019ApJ...876...27Y}. 
We have included free-free
and self-synchrotron absorption for completeness, but these effects are not influential to our results. 
We focus on the non-thermal emission from particle acceleration at the FS and ignore any possible contribution from the reverse shock (RS). 
For the SN ejecta, the parameters are chosen to represent a canonical thermonuclear explosion model with a near-Chandrasekhar mass white dwarf, i.e., 1.4~M$_\odot$ for the ejecta mass and an explosion energy of 1.0$\times10^{51}\ \mathrm{erg}$. We adopt an exponential density profile\footnote{We note that the asymmetries in Tycho's SN ejecta itself can possibly affect the evolution of its remnant \citep{Ferrand_2019,2022ApJ...937..121M}, but we ignore the mutli-dimensional effects here for simplicity.} for the initial ejecta distribution. \citep{1998ApJ...497..807D}. 

\section{Results and Discussion}\label{sec:res}

\begin{figure}[hpt]
    \epsscale{1.15}
    \plotone{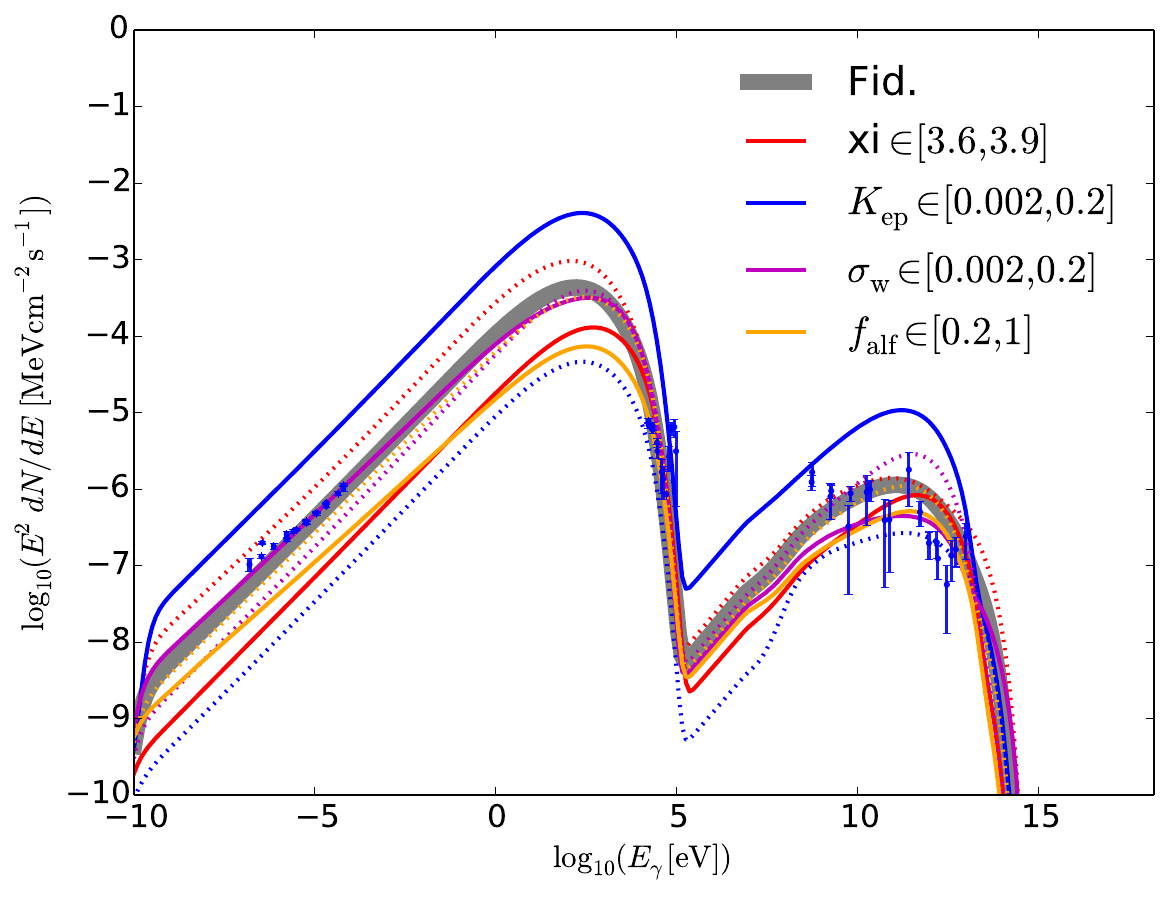}
    \caption{Broadband non-thermal emission spectrum which has the solid angle of $4\pi$ when varying and calibrating parameters for particle acceleration using the dynamical parameters of Region 13. The data points with error bars show the observed fluxes from radio observations \citep{2006AA...457.1081K}, X-ray data from \textit{Swift}/BAT \citep{2014ApJ...797L...6T}, GeV gamma-ray data from \textit{Fermi}-LAT \citep{2017ApJ...836...23A,2012ApJ...744L...2G}, and TeV gamma-ray data from \textit{VERITAS}\citep{2017ApJ...836...23A,2011ApJ...730L..20A}. }
    \label{fig:res-0sedcalib}
\end{figure}

\begin{deluxetable*}{l||cccc|ccc|cc}
\centering
\label{tab:model-params}
\tablecolumns{3}
\tablewidth{15cm}
\tablecaption{Input parameters for particle acceleration and the ambient environment for each model. The rightmost two columns summarize the compatibility of the models with the proper motion data and the broadband spectral data.} 
\tablehead{
Model&$\chi_\mathrm{inj}$&$K_\mathrm{ep}$&$\sigma_\mathrm{w}$&$f_\mathrm{alf}$&$\dot{M}/(4\pi V_\mathrm{w})$&$\rho_\mathrm{outer}$\tablenotemark{*}&$B_0$&  Shock exp.&Emission\\
&-&-&-&-&$\mathrm{g\ cm^{-1}}$&$\mathrm{g\ cm^{-3}}$&$\mathrm{\mu G}$  &(2003-2015)&(2010s)\\
Range{**}&3.9&$2.0\times10^{-3}$&$2.0\times10^{-3}$&1& -&-&- &-&-\\
&3.6&$2.0\times10^{-1}$&$2.0\times10^{-1}$&0.1& -&-& -&-&-
}
\startdata
A\tablenotemark{a}& 3.75&$2.0\times10^{-2}$&$2.0\times10^{-2}$&0.1& $\sim10^{13}$&$\sim10^{-22}$&-  &$\surd$&$\times$($\gamma$-ray)\\
B& &&\hspace{-15.0mm}(Same as A)&& $\sim10^{13}$&$\sim4\times10^{-23}$&-  &$\surd$&$\times$($\gamma$-ray)\\
C\tablenotemark{$\dagger$,b}& 3.6&$8.0\times10^{-3}$&-&0.7& -&$0.3 m_\mathrm{
p}$&5.0  &$\times$&$\surd$\\
\enddata
\tablenotetext{*}{The gas density at the outer boundary of the simulation box. }
\tablenotetext{**}{The upper limit (upper row) and the lower limit (lower row) range of parameters for fitting Models A and B.}
\tablenotetext{\dagger}{Model C is similar to the best-fit model in \citet{Slane2014} except that we have chosen a different value for $f_\mathrm{alf}$ here to obtain the best spectral fit.}
\tablenotemark{a}{The distance to Tycho's SNR is $D = 3.5$~kpc in 
the geometric center case (fiducial) and $3.7$~kpc in the pressure center case in accordance to \citetalias{2024ApJ...961...32K}.}\\
\tablenotemark{b}{The distance is $D=3.18$~kpc according to \citet{Slane2014}.}
\end{deluxetable*}

\begin{figure*}[ht]
    \epsscale{1.15}
    \gridline{\fig{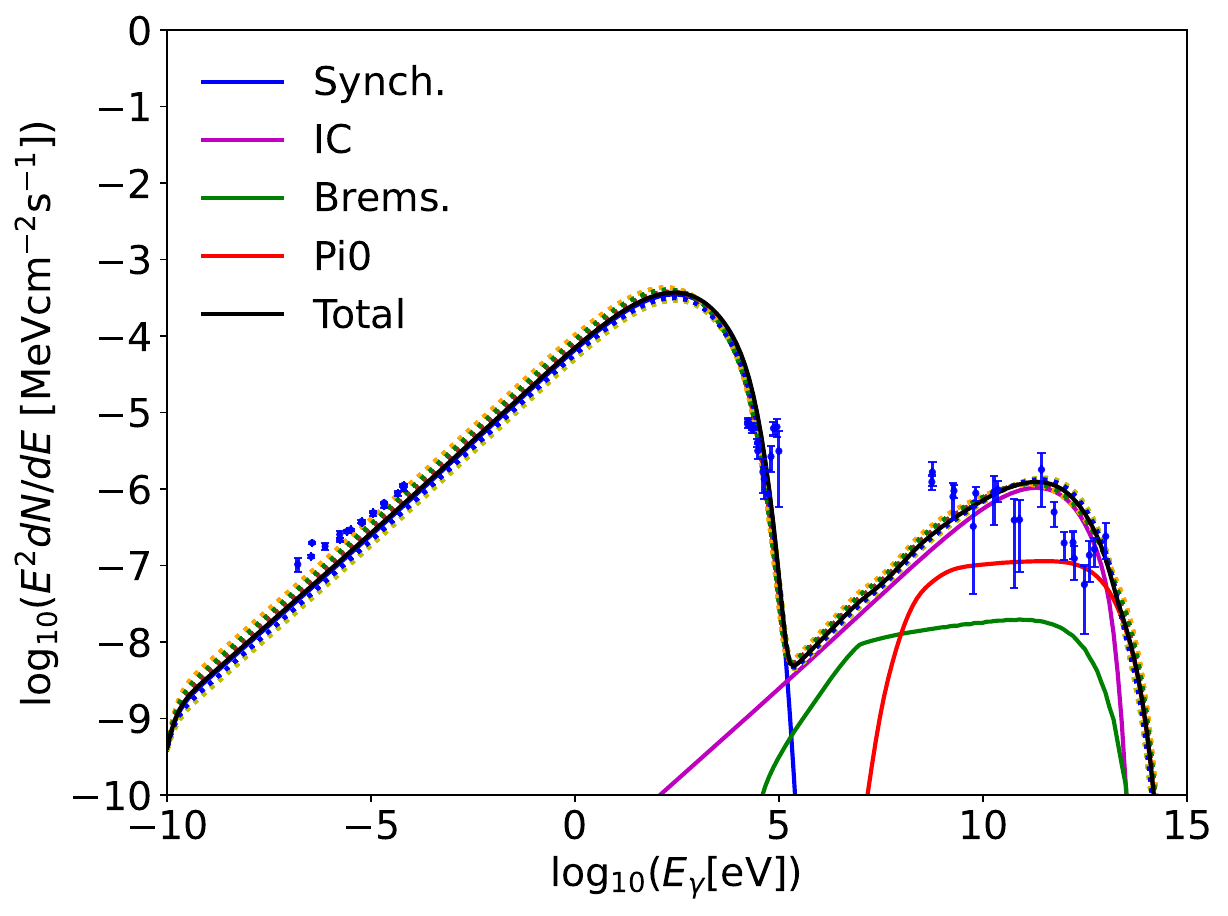}{0.5\textwidth}{(a) Azimuthally integrated spectrum}
    \fig{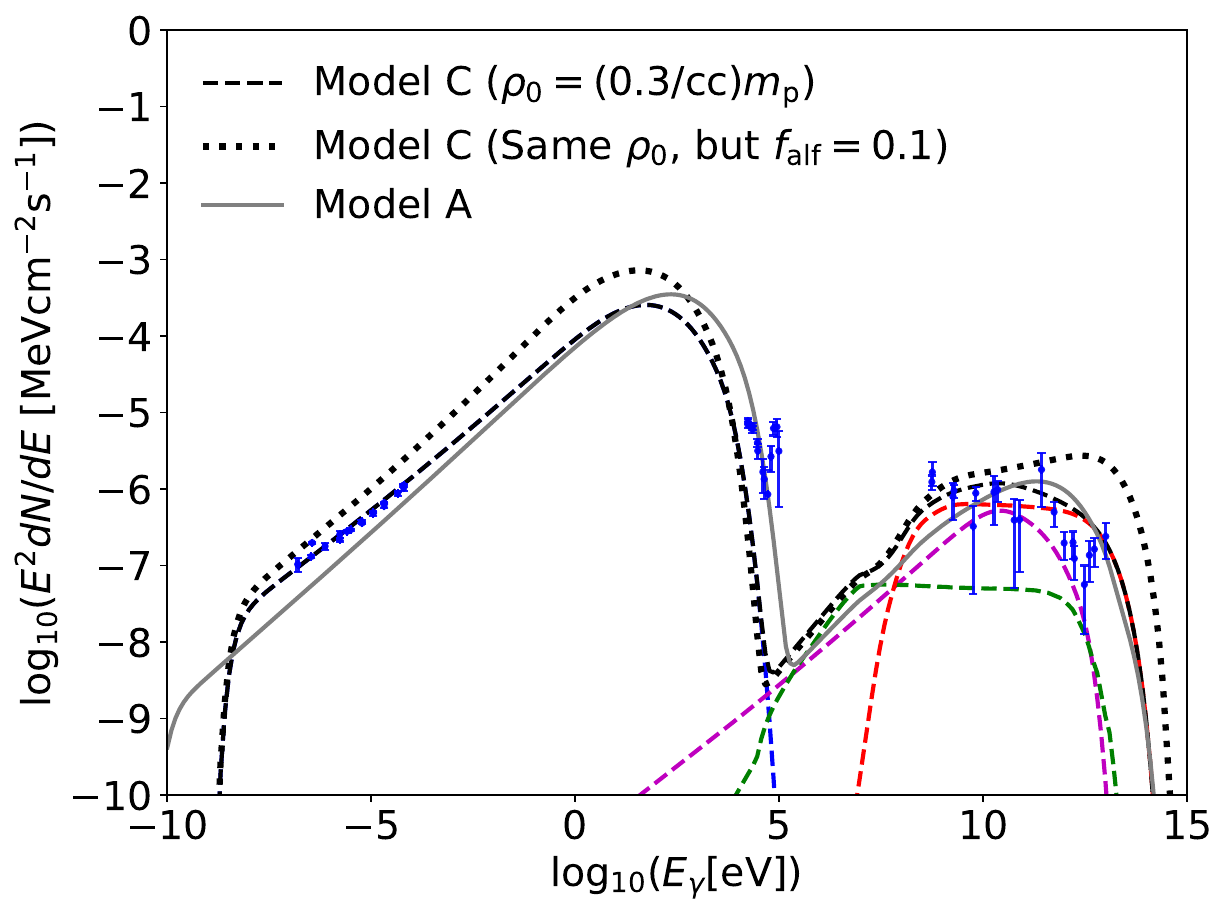}{0.5\textwidth}{(b) Comparison with Model C (uniform medium)}}
    \gridline{\fig{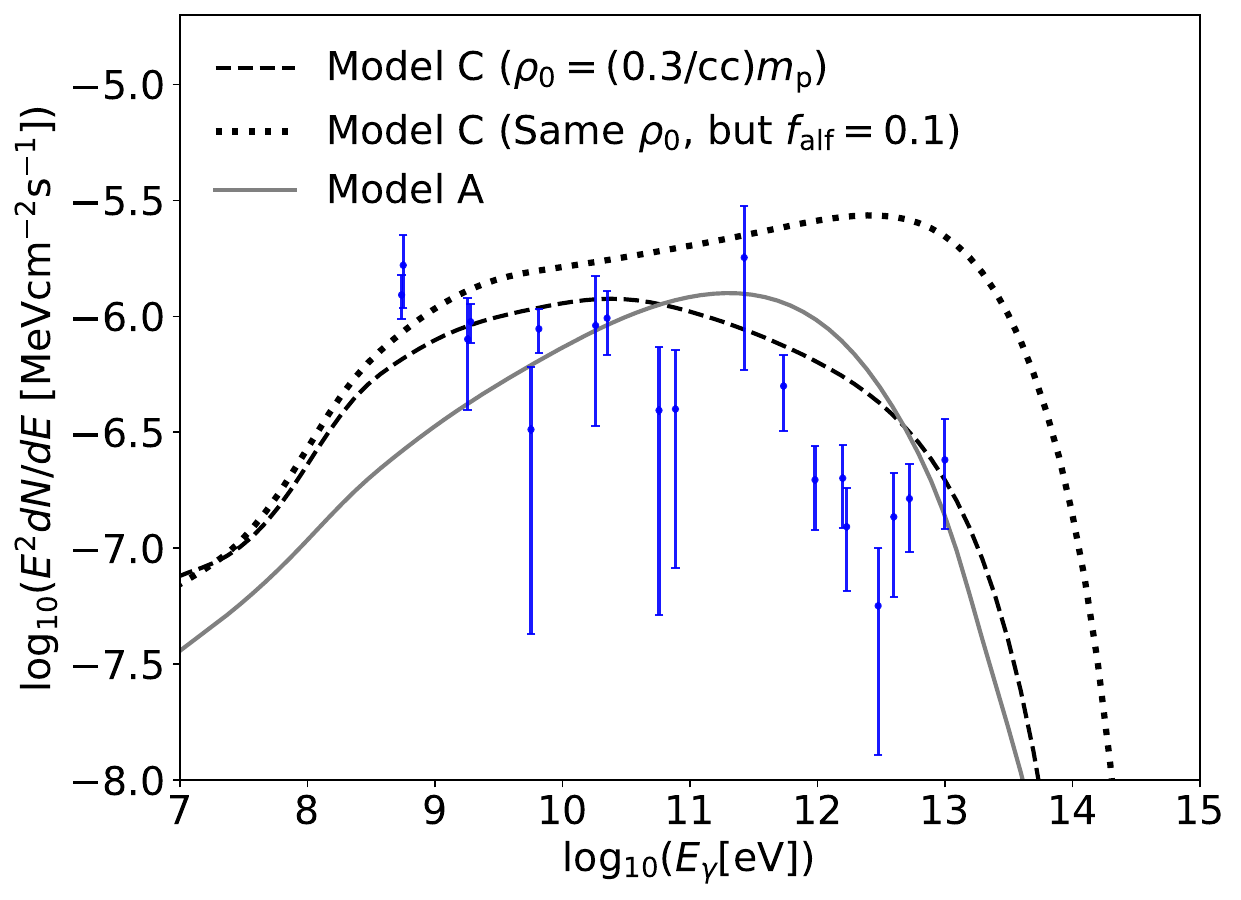}{0.5\textwidth}{(c) Comparison with Model C, zoomed-in}\fig{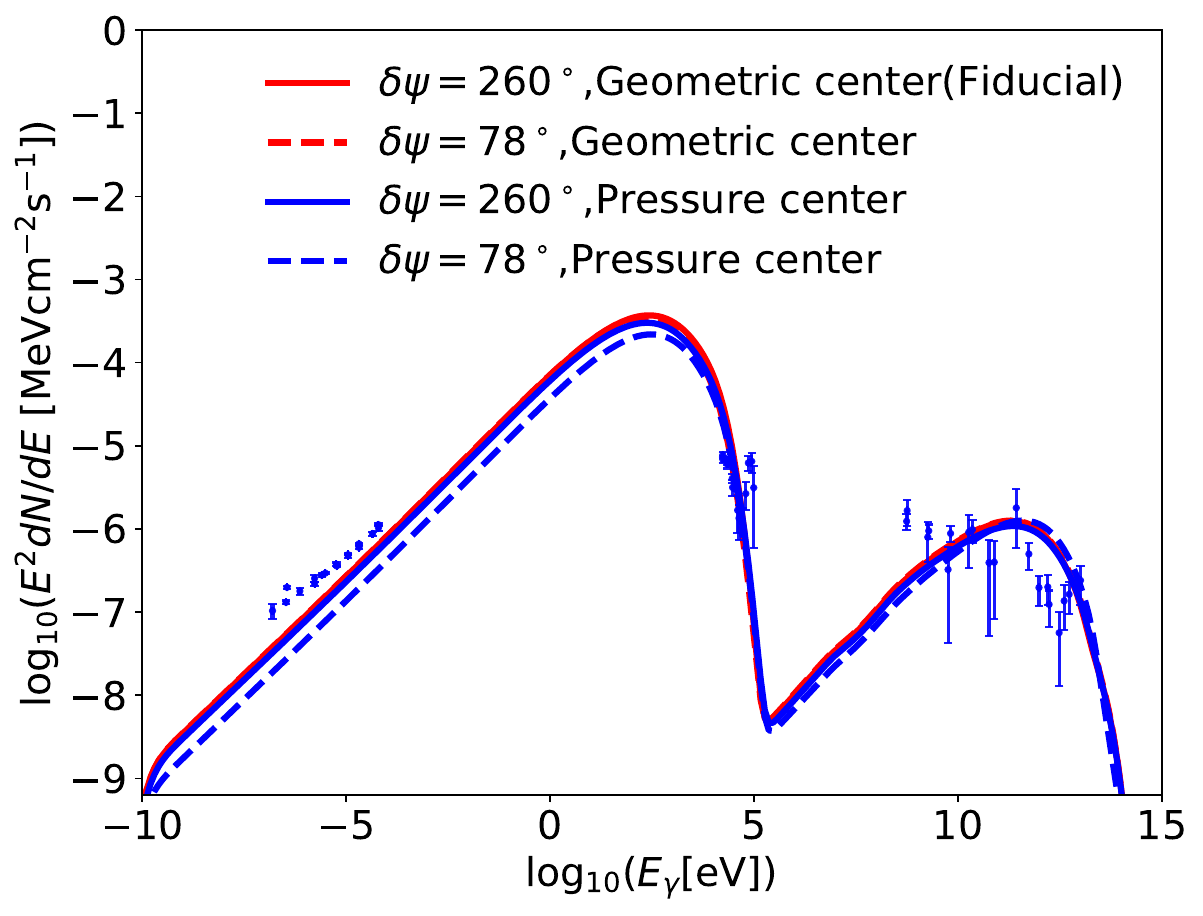}{0.5\textwidth}{(d) Dependence on explosion center}}
    \caption{
    Spectral models of the broadband non-thermal emission of Tycho at an age of 440~yrs (or in the year of 2012). 
    The data points are same as in Figure~\ref{fig:res-0sedcalib}. Panel (a): The black solid line shows the total azimuthally-integrated (averaged) spectrum which can be decomposed into four non-thermal components, synchrotron (blue solid), bremsstrahlung (green solid), inverse compton scattering (magenta solid) and $\pi^0$ decay (red solid). To show the variation against the azimuthal angle, spectra calculated from the best-fit models for each of the 13 different azimuthal regions are also shown by the dotted lines in different colors. 
    Panel (b): Comparing the azimuthally-integrated spectrum from Model A (solid line) with Model C with a uniform ambient medium (dashed line and dotted line), decomposed into each emission component using the same color as in Panel (a). The main difference between the dashed line and dotted line is the efficiency of magnetic field amplification. 
    Panel (c): Same as Panel~(b) but zoomed into the gamma-ray band to better visualize the gamma-ray spectral index. 
    Panel (d): Model A at 2012 but for different offset azimuth ($\delta\psi=78.4^\circ$ and $260.3^\circ$ in solid and dashed lines respectively) and different explosion center assumed (geometric center in red and pressure center in blue). All spectra shown are calculated for the year of 2012. 
    }%
    \label{fig:res-1sed}
\end{figure*}

Since the differences between the model spectra characterizing the different azimuthal regions are found to be not significant, it is a good approximation to calibrate the parameter set for particle acceleration based on the result for a representative region such as Region 13. Fig.~\ref{fig:res-0sedcalib} demonstrates how each of the parameter affects the resulting emission spectrum from the entire shell of the SNR with the dynamical paramters same as Region 13\footnote{Our modeled emission flux is for the whole remnant calculated by an angular integration of the fluxes from each of the 13 regions. The calibration of acceleration parameters was done based on Region 13, which was performed before the angular integration over the 13 regions. This can be justified since the difference of the modeled spectra of different regions turns out to be insignificant (Fig.~\ref{fig:res-1sed}a). We calibrate more easily comparing between the data and the model based on Region 13 for fitting. }.  
Most data in the radio band have been taken in the 1970s which is a different timing than the data in other wavebands. We see an increasing trend of the flux with age in Models A and B and the opposite for model C, but the rate of change in flux is found to be small in either cases before 2012; the influence from this timing difference is thus negligible, at least until the shock has started to penetrate deeply into the outer dense shell.
We set the fiducial model parameters for particle acceleration identical for all the 13 regions and for Models A and B which can reproduce the observed flux from Tycho in 2012, with a goodness-of-fit of $\chi^2/\mathrm{dof}\sim144,\ \mathrm{dof}=48$: $(\chi_\mathrm{inj},K_\mathrm{ep},\sigma_\mathrm{w}, f_\mathrm{alf})=(3.75, 2.0\times10^{-2},2.0\times10^{-2},0.1)$ (see Table~\ref{tab:model-params}). The range of each parameter is shown as in Table~\ref{tab:model-params}, and compare the fiducial value model with the lower and upper limit values for order estimate. The critical parameter to our results is $K_\mathrm{ep}$, which affects the normalization of the overall spectrum, and $f_\mathrm{alf}$ changes the gamma-ray spectral index. 
The parameter set for Model C is largely based on the parameter set of ``Model A'' in \citet{Slane2014} as described in Table~\ref{tab:model-params}, but uses a higher $K_\mathrm{ep}$.

Figure~\ref{fig:res-1sed}a shows the broadband spectrum resulting from Model A, which shows a good match to the observed fluxes in all wavelengths in 2012. An exception is the gamma-ray regime, where we cannot reproduce the softness of the spectrum found at energies $\sim$~TeV in this model. In the case of Model C as shown in Figures~\ref{fig:res-1sed}b and ~\ref{fig:res-1sed}c, the spectral slope in gamma-ray is reproduced, as is explained by the hadronic contribution which is in line with a few previous studies \citep[e.g.,][]{Slane2014,2019ApJ...876...27Y}. The model dependence on $f_\mathrm{alf}$ in Model C is shown by the difference between the dashed line and the dotted line in Figure~\ref{fig:res-1sed}c, from which we see a better fit for a $f_\mathrm{alf}$ value similar to that in \citet{Slane2014}. 
For Model A, changing $f_\mathrm{alf}$ from its best fit value results in even a larger deviation of the model spectrum from the observation data as shown in Figure~\ref{fig:res-0sedcalib}. 

Despite the large error bars, the observed spectral index appears to be softer than 2.0, which favors Model C to our new Model A from a wind-shell environment. This difference in spectral hardness comes from the very different upstream gas density at an age of around 400~yr for these models, which results in a hadronic-dominated emission in Model C and a leptonic-dominated emission in Models A and B. The former has the shock interacting with a relatively dense ISM gas whereas the latter finds its shock propagating in the rarefied outer part of the wind bubble (as one can see in Figures~\ref{fig:denden-init} and \ref{fig:res-2acc}a). 
The challenge of reproducing the gamma-rays promotes further investigation into a new CSM environment model for Tycho, such as one with clumpy clouds as discussed in Section~\ref{sec:disc}. As shown in Figure~\ref{fig:res-1sed}d, the exact 3-D geometry of the wind cavity and the denser region engulfing Tycho will play some roles in the spatial variation and time evolution of the shock dynamics and the overall emission spectrum as well \citepalias[see also][]{2024ApJ...961...32K}.

\begin{figure*}[ht]
    \epsscale{1.15}
    \gridline{\fig{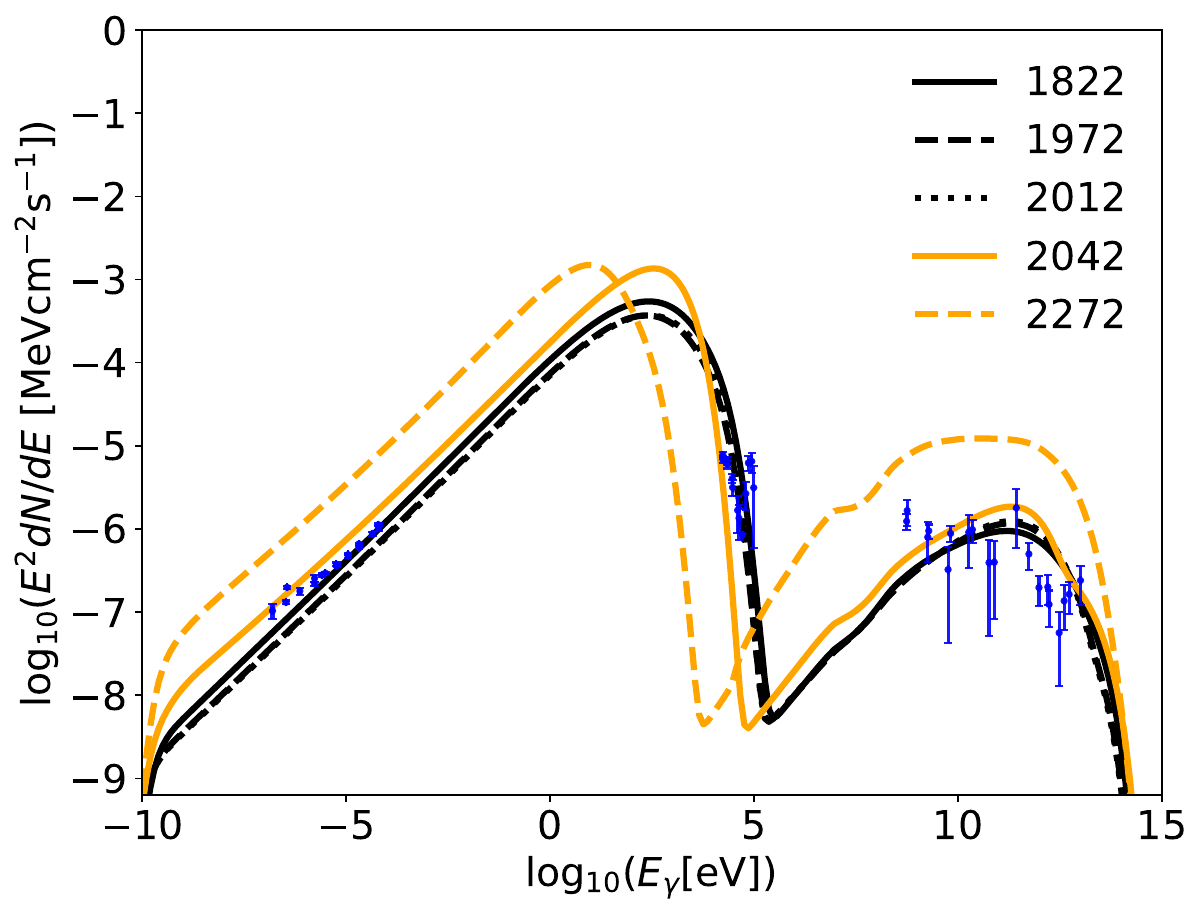}{0.5\textwidth}{(a) Evolution of Model A over the period of 1820 -- 2270}\fig{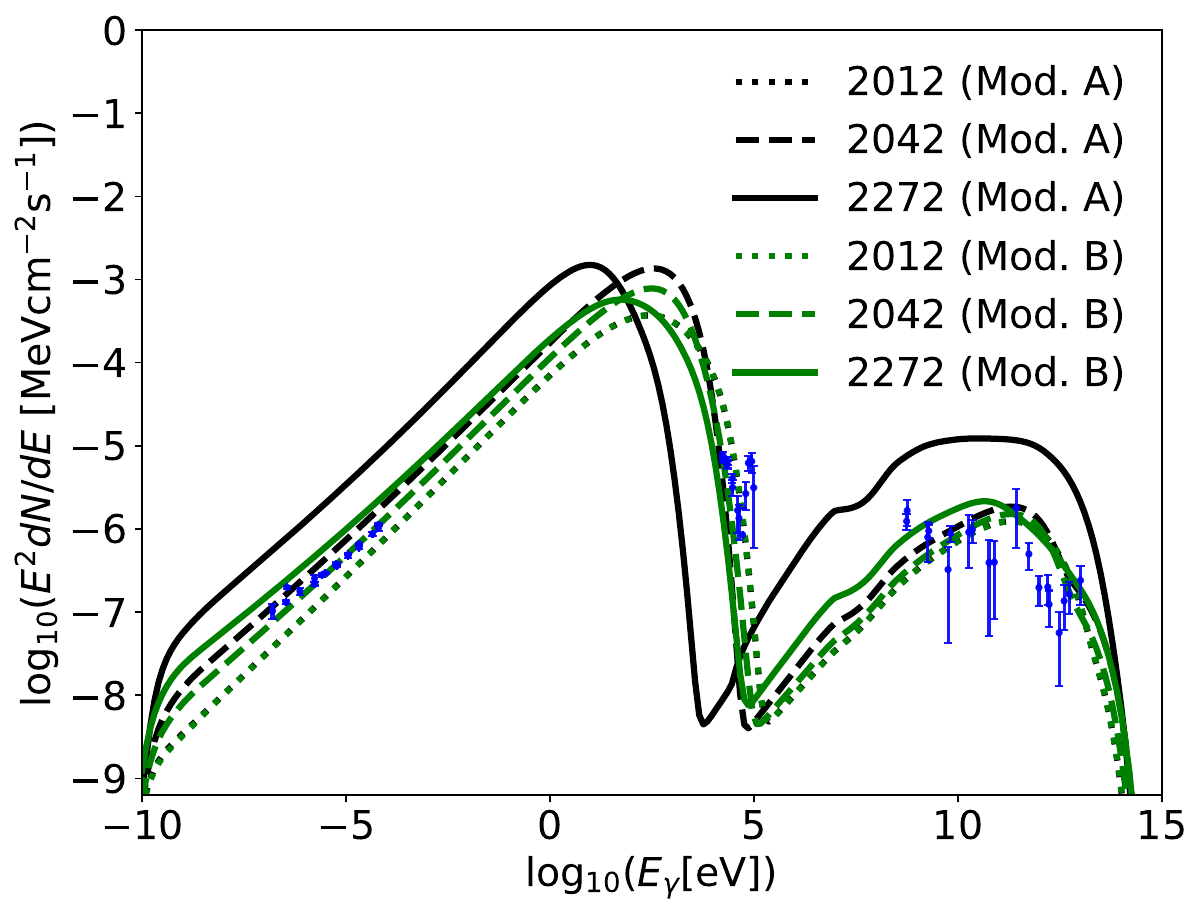}{0.5\textwidth}{(b) Comparison with Model B (2010 -- 2270)}}
    \caption{Predictions on the broadband SED including epochs of 2042, 2272 when the shock has already been interacting with the cloud. The data points are the same as in Figure~\ref{fig:res-0sedcalib}. Panel (a): The calculated spectrum of Model A in 1822, 1972, 2012, 2042 and 2272 which correspond to the black dotted, dashed, solid lines, orange solid, dashed lines. Panel (b): Evolution over the period of 2010 -- 2270 compared with Model B. Model A in 2012, 2042 and 2272 are shown by the black dotted, dashed and solid lines respectively, and Model B in the same years are shown in green with the same line types.}
    \label{fig:res-3sedfut}
\end{figure*}

\begin{figure*}[ht]
    \epsscale{1.15}
    \plotone{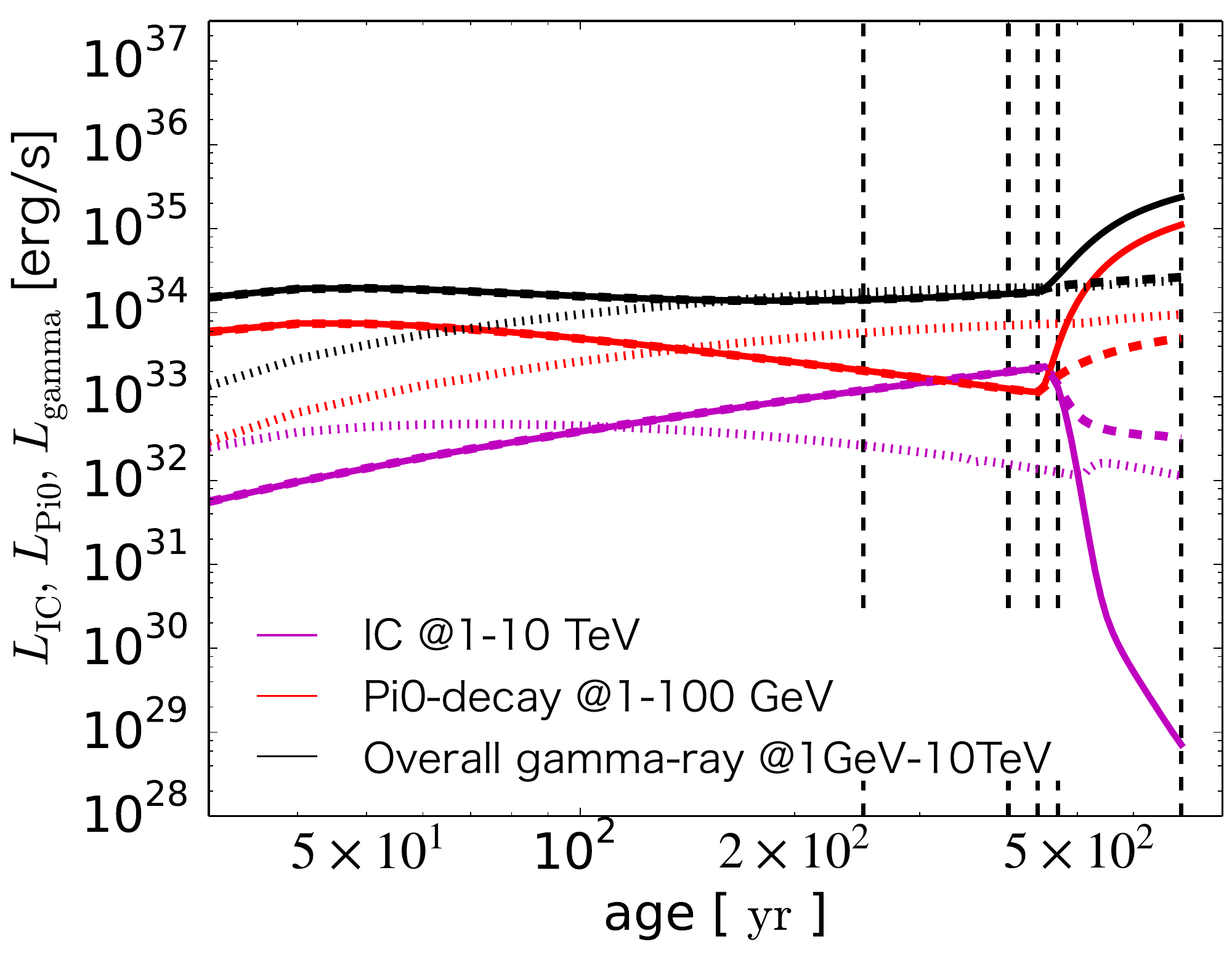}
    \caption{The light curve of the leptonic/hadronic gamma-rays for Model A (solid lines), Model B (dashed lines) and Model C (dotted lines). For our new Models A and B, the leptonic emission via IC is found to be brightening gradually before the interaction with clouds, and after the interaction began it delines rapidly. The hadronic component via $\pi^0$-decay on the other hand shows an opposite trend of evolution. The vertical lines show the years 1822, 1972, 2012, 2042 and 2272 for reference. }
    \label{fig:res-2lc}
\end{figure*}

\begin{figure*}[ht]
    \epsscale{1.15}
    \gridline{\fig{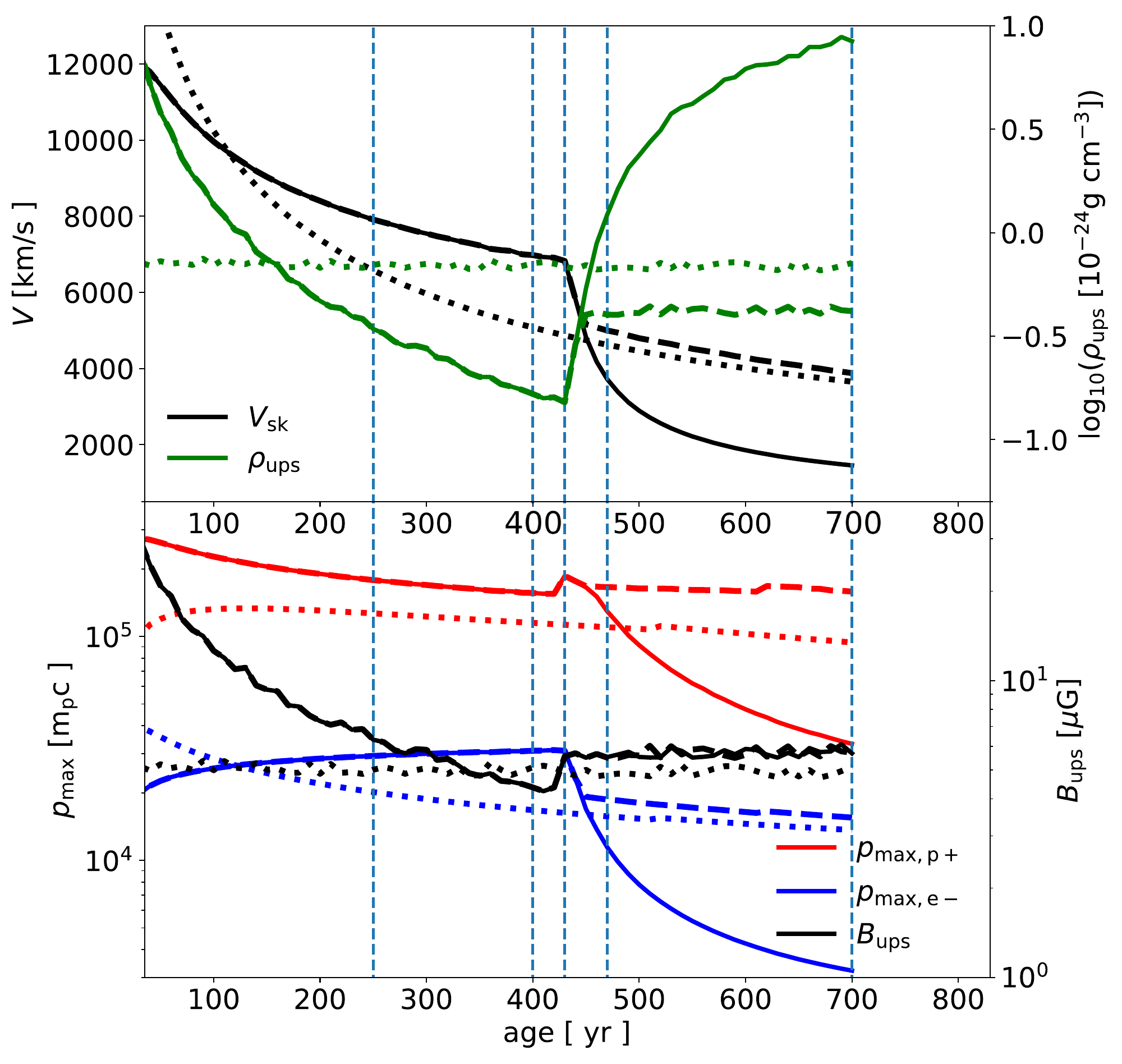}{0.5\textwidth}{(a) Time evolution of shock speed, upstream states and $p_\mathrm{max}$ of accelerated particles}\fig{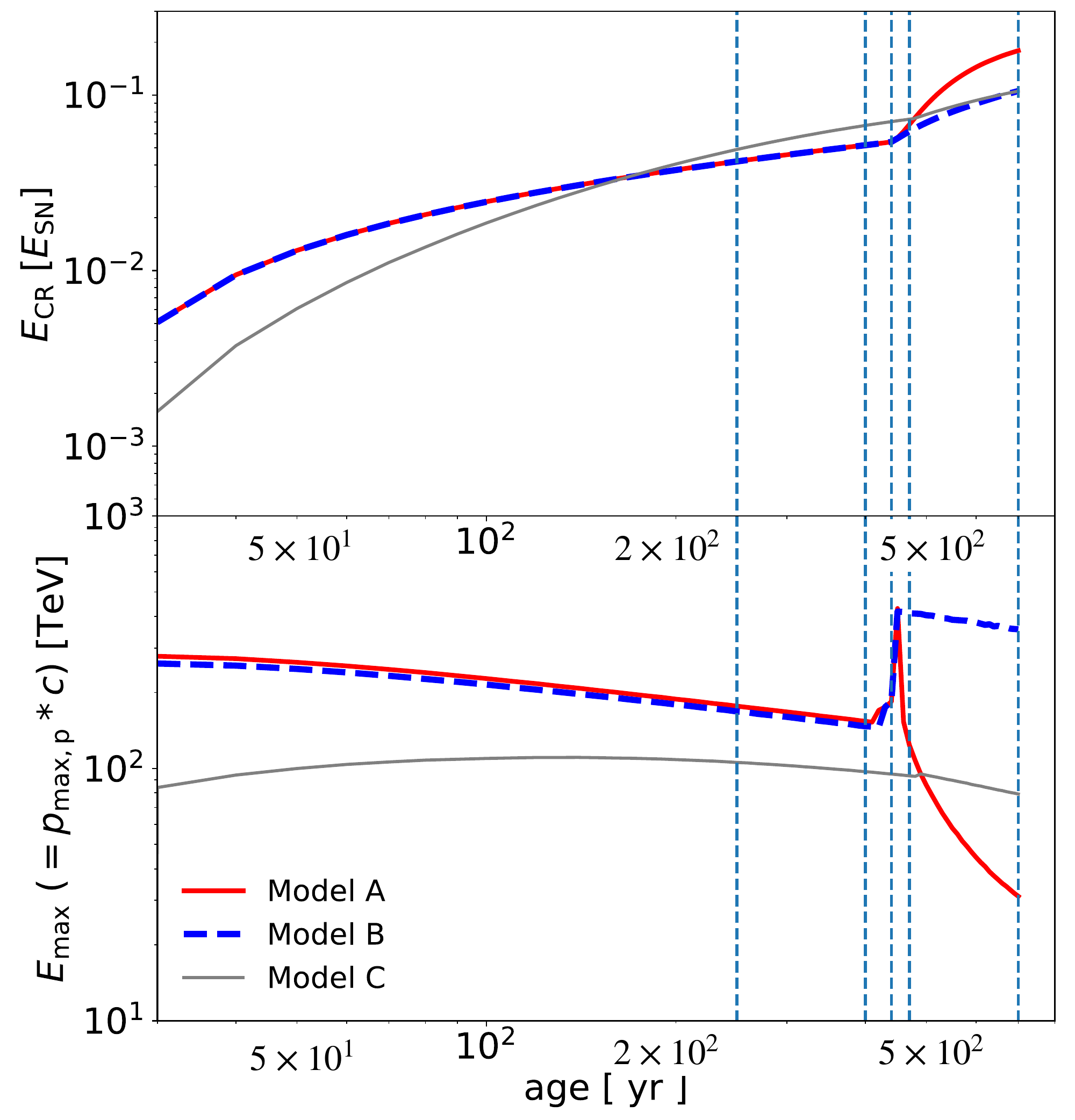}{0.5\textwidth}{(b) Total energy and $p_\mathrm{max}$ of accelerated protons versus age}}
    \caption{Time evolution of a few selected key quantities predicted by our three models. 
    The vertical lines again show the year 1822, 1972, 2012, 2042 and 2272 for reference. Panel (a): Time evolution of the shown properties in Region 2 from Model A (solid lines), Model B (dashed lines) and Model C (dotted lines). Upper panel: Time evolution of the shock velocity and the density just upstream of the shock. Lower panel: Time evolution of the maximum momenta and the upstream magnetic field strength. The red and blue lines indicate the maximum momentum of protons and electrons respectively. Panel (b): Time evolution of particle acceleration history for Model A, B and C,  as shown by the red solid, blue dashed, grey solid lines, respectively. Upper panel: The total kinetic energy in the accelerated CRs $E_\mathrm{CR}$ normalised by the kinetic explosion energy $E_\mathrm{SN}$, integrated over all the azimuthal regions . Lower panel: The CR proton maximum energies $E_\mathrm{max}$ (from the maximum value among the 13 regions). The vertical lines are the same as in Panel (a). }
    \label{fig:res-2acc}
\end{figure*}

\begin{figure}[ht]
    \epsscale{1.15}
    \plotone{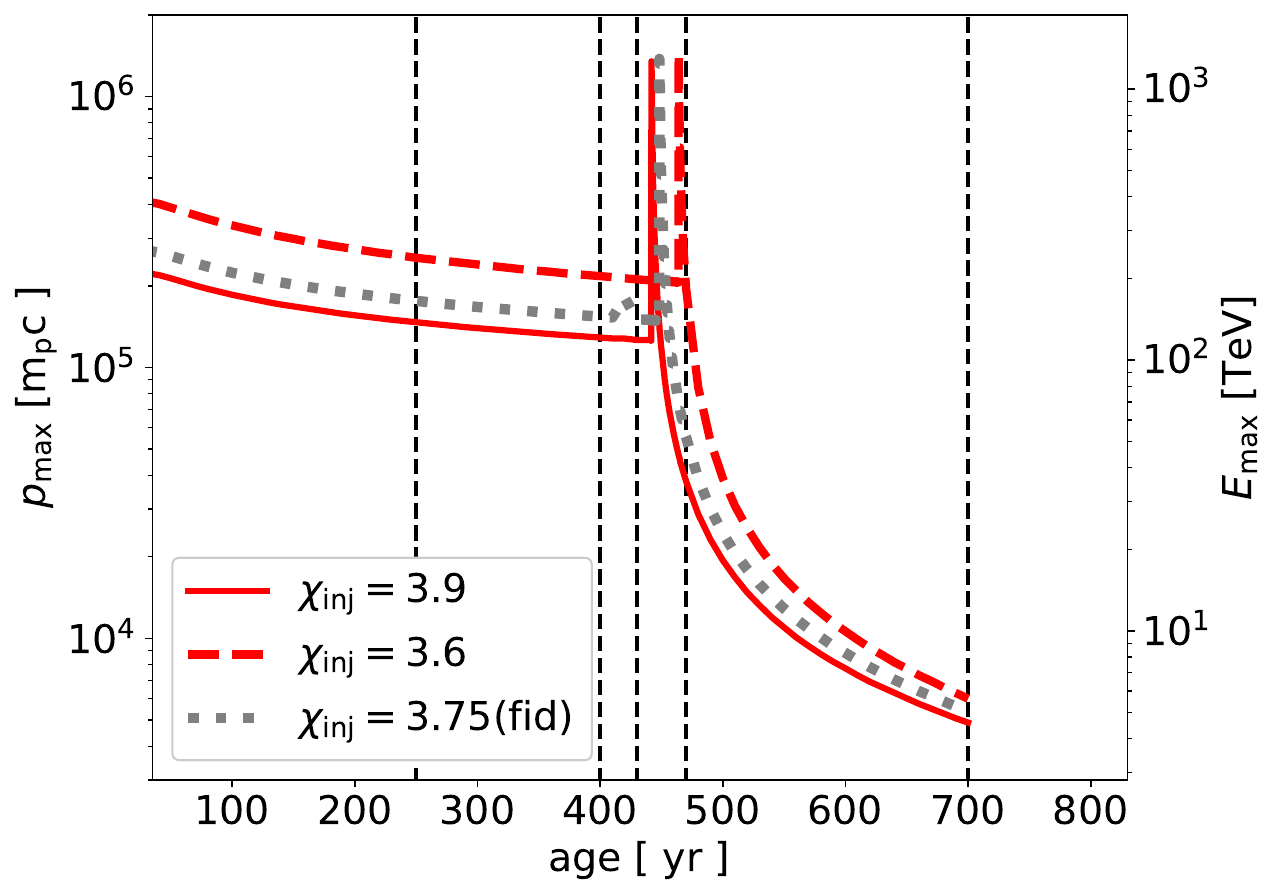}
    \caption{Time evolution of maximum momentum of protons in Model A and other models with different acceleration efficiencies $\chi_\mathrm{inj}$ shown with a smaller timestep. Time evolution of particle acceleration physics from the region with the highest $p_\mathrm{max}$ reached (Region 7) is a fiducial model ($\chi_\mathrm{inj}=3.75$) is expressed in grey dotted line as well as different models with different acceleration efficiencies ($\chi_\mathrm{inj}=3.6,3.9$) in red dashed and red solid line. The vertical lines show year 1822, 1972, 2012, 2042, 2272.}%
    \label{fig:res-3acc}
\end{figure}

\if0\begin{figure*}[ht]
    \epsscale{1.15}
    \plotone{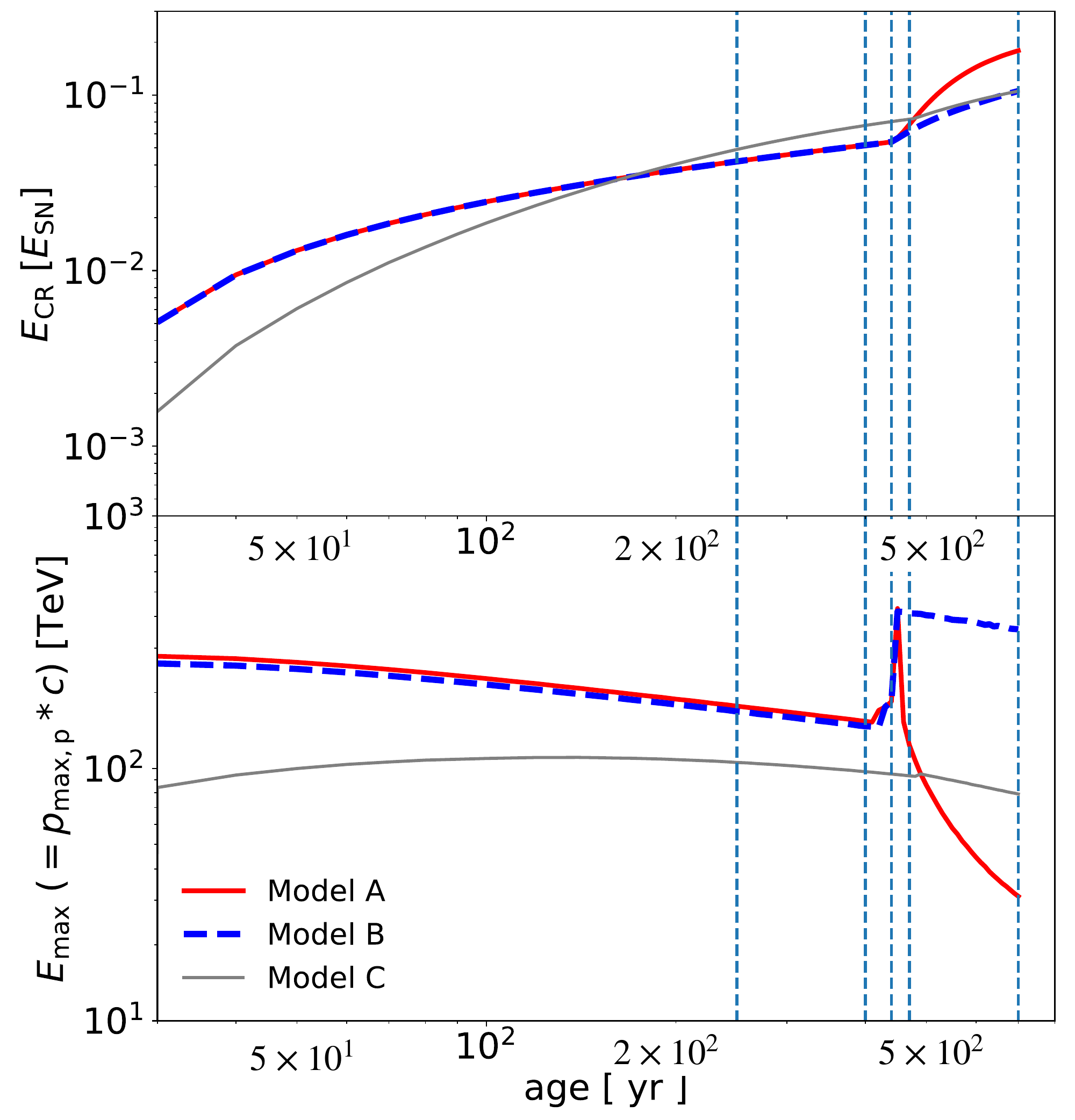}
    \caption{Time evolution of particle acceleration physics over the whole regions in Model A, B and C (red solid, blue dashed, grey solid lines, respectively). Upper panel: The total kinetic energy in the accelerated CRs $E_\mathrm{CR}$ normalised by the kinetic explosion energy $E_\mathrm{SN}$. Lower panel: The maximum value in the CR maximum energies $E_\mathrm{max}$ in different 13 regions. The vertical lines show year 1822, 1972, 2012, 2042, 2272.}%
\end{figure*}\fi

As for the time evolution of the SED, the overall spectrum decreases in flux monotonically due to an adiabatic loss from the expansion of the SNR until the interaction with the outer dense shell begins (see Figure~\ref{fig:res-3sedfut}a). Up until 2012 (i.e., soon after the beginning of the interaction with dense clouds at $\sim$~2007), the hadronic contribution in the gamma-ray emission via $\pi^0$-decay is decreasing with time while the leptonic contribution via IC shows the opposite as we can see in Figure~\ref{fig:res-2lc}. The interaction with the dense shell causes a reversal of the trend where the pion-decay gamma-rays experience a sharp boost from the enhanced upstream gas density whereas the IC component declines rapidly from a fast synchrotron loss. 

The time evolution concerning the upstream conditions at the shock and particle acceleration are shown in Figure~\ref{fig:res-2acc}a. As confirmed in \citetalias{2024ApJ...961...32K}, the shock velocity shown in the upper panel of Figure~\ref{fig:res-2acc}a undergoes a substantial deceleration which reproduces the observation by \citetalias{2021ApJ...906L...3T}. 
This reflects accordingly in the time dependence of the upstream gas density and the magnetic field strength as shown in Figure~\ref{fig:res-2acc}a. Both increase at almost the same timing as the shock deceleration. The maximum momenta of the accelerated protons and electrons  $p_\mathrm{max,p/e}$ are shown in the lower panel of Figure~\ref{fig:res-2acc}a. We see that $p_\mathrm{max,e}$ undergoes a gradual increase inside the wind bubble, while decreases substantially after the interaction with the dense clouds has begun as synchrotron loss becomes important. $p_\mathrm{max,p}$ undergoes a substantial decline either, in the phase of the shock deceleration. The slight increase in $p_\mathrm{max,p}$ found before the fast decay is due to a temporary boost of magnetic field amplification from the shock-cloud interaction. These trends are much less pronounced in Model B due to the flatness of the gas density assumed for the environment beyond $R_{2015}$. 

We have also calculated the history of cosmic-ray acceleration in these models, as represented by the time evolution of $E_\mathrm{CR}(t)$ and $E_\mathrm{max}(t)$. The upper panel of Figure~\ref{fig:res-2acc}b shows the time evolution of the total kinetic energy $E_\mathrm{CR}$ of the accelerated CRs integrated over the whole remnant lifetime, as is normalised by the explosion energy $E_\mathrm{SN}$. The values in Models A and B in 2012 are found to be moderate at around $4\times10^{-2}$, while it is $\sim7\times 10^{-2}$ in Model C. After 2010's, it increases to exceed $10^{-1}$ until 2170's (600~yr) in Model A. 
$E_\mathrm{max}=\max\{p_\mathrm{max,p}c\}_\mathrm{region}$ is the maximum value of the maximum proton energy among the 13 regions, which is shown in the lower panel of Figure~\ref{fig:res-2acc}b. Model C predicts the value of $\sim10^{14}$~eV in 1670's (an age of around 100~yr), which then decreases monotonically until the 2170's; Models A and B, on the other hand, predict the maximum energy of around $2\times10^{14}$~eV in 1670's, which then decreases until 1970's, followed by a rapid rise to a peak and the fall in 2040's due to the enhanced magnetic field and the subsequent rapid shock deceleration in the dense clouds. 
We see that a $E_\mathrm{CR}/E_\mathrm{SN}\sim10\%$ is reached for all Model A, B, and C 
which agrees with the scenario of SNRs being the main origin of Galactic CRs from the energetics point of view \citep[e.g.,][]{1987PhR...154....1B,LEN2012,2013ApJ...777..148G, 2022ApJ...936...26K}. 
As shown in Figure~\ref{fig:res-3acc}, the evolutionary trend will qualitatively be similar over different CR injection efficiencies $\chi_\mathrm{inj}$, despite the moderate change in the abolute value and the timing that the maximum energy reaches its peak when the FS hits the cavity wall,
due to a modification to the shock dynamics and magnetic field strength.

As is the case for many other time evolution and particle acceleration models of SNRs, $E_\mathrm{max}\sim3\times10^{15}$~eV, corresponding to the ``knee'' feature in the observed Galactic CR spectrum, is not explained by the present models for Tycho.
There have been claims that the knee energy is reached in some simulations \citep[e.g.,][]{2019ApJ...872...46I,2021ApJ...922....7I,2023ApJ...958....3D}, but it has been realized only in a relatively short period of time, which is too short for SNRs to be the main source of such high-energy particles. This is an issue that requires further investigation. 

Our hydro models predicts the spectral evolution after 2015 (Figure~\ref{fig:res-3sedfut}a,b), which can be compared to upcoming future observations. In Model A after 2015, i.e., after the interaction with the MC (Figure~\ref{fig:res-3sedfut}a), the spectral evolution is characterized by the combination of a fast synchrotron loss for the electrons, the rapid shock deceleration, and the increased target gas density in the dense upstream region. We can see a sharp flux increase from $\sim$2010's due to the shock entering into the dense region. Model B does not involve a large increase in the upstream density, keeping a moderate flux level (Figure~\ref{fig:res-3sedfut}b).
These trends are also seen in the gamma-ray light curves as shown in Figure~\ref{fig:res-2lc}. Moreover, we can see in Figure~\ref{fig:res-2lc} that in 2010's the hadronic emission (i.e., $\pi^0$ decay) is increasing while the leptonic emission (i.e., IC) is decreasing; the former thus exceeds the later. Observational data to compare with our prediction after 2015 are not yet available \citep[e.g.,][]{2022ApJ...930..151X}. The density distribution beyond $R_{2015}$ can/will be addressed once such data (both dynamical and spectroscopic) become available. 


\section{Possible multi-dimensional effects}\label{sec:disc}

In this section, we discuss a few issues on how multi-dimensional effects, which are not included in this work, would affect our conclusions. Clumpy structure in the CSM is suggested to generate strong turbulence downstream of the shock \citep[e.g.,][]{2012ApJ...744...71I}, which may well affect the formation of the non-thermal spectrum. For example, it is expected that such configuration will enhance the particle acceleration, due to the excitation of magnetic turbulence and reflected shocks generated by the dense upstream material. We also expect an enhancement of the $\pi^0$-decay gamma-rays, due to the dense clumpy medium acting as a target for the pion production. These effects are not considered here because the current one-dimensional work essentially averages out such structures. The enhancement of the hadronic gamma-ray emission through the clumpy medium may help reconcile the difficulty in Models A and B in reproducing the steepness of the gamma-ray emission, as Figure~\ref{fig:res-1sed}c suggests. 

Multi-dimensional simulations can possibly reconcile another discrepancy between our model and observation. From the morphology of Tycho (e.g., in the radio band), Rayleigh-Taylor fingers around the shocked material have been confirmed in some regions \citep{1998AA...334.1060V}. These features can protrude the FS and thus change the shock structure, which may lead to an azimuthal variation of the FS radii \citep{2023AA...672A.194R}. This is not yet included in this work. 

Moreover, we did not consider the true three-dimensional structure of Tycho's environment by assuming an axisymmetry as prescribed in Eq.~\ref{eq:sum-up}; our hydro models have been mainly constrained by proper motion measurements along the plane of the sky, only showing near the rim of the shock in the object. Taking into account observational data constraining the Doppler motion along different line-of-sight directions which also showed the substantial shock deceleration \citep[see e.g.,][]{2022ApJ...937..121M,2024ApJ...962..159U,2024arXiv240417296G} will provide a more realistic picture for the environmental properties of the Tycho's SNR in 3-D. 

An additional caveat is that our model setup for the density profile of Tycho's environment is based on a simplified toy model which aims at qualitatively representing the basic features of a wind-cavity CSM structure; this work aims at a first qualitative estimate of the non-thermal emission from the Tycho's SNR taking into account the latest observational information on its environment through the proper motion measurements.
More accurate and detailed models can be in principle constructed by 3-D hydro simulations with realistic parameters. We plan to address this issue in a future follow-up paper.

Another issue yet to be addressed is the consistency of the new models with the thermal emission properties. Next-generation telescopes such as the recently launched \textit{XRISM} will provided new insight for the origin of this important type Ia SNR through high resolution X-ray spectroscopy; the joint modeling of the thermal and non-thermal emissions as well as the dynamical evolution will be an important step in the future.  

\section{Summary}\label{sec:sum}
Although a homogeneous environment for the Tycho's SNR had been suggested for many years \citep[e.g.,][]{2005ApJ...634..376W,Slane2014}, a recent re-analysis of the shock-expansion dynamics from proper motion observations using \textit{Chandra} by \citetalias{2021ApJ...906L...3T} has revealed a recent substantial deceleration of its shock, inferring the existence of dense clouds outside of a low-density CSM \citep[see also][]{2016ApJ...826...34Z}. \citetalias{2024ApJ...961...32K} showed that such an environment with the wind-MC density structure can explain the shock-expansion data presented by \citetalias{2021ApJ...906L...3T}.
The present paper extends the work further to the non-thermal emission properties, to test whether the updated environment model is consistent with the observed spectrum. 
%

Our results show that the observed non-thermal spectrum can be largely reproduced by the wind-MC environment. One drawback is that such an environment has a difficulty in reproducing the gamma-ray data; the predicted gamma-ray spectrum is harder than observed. This is mainly due to the enhanced importance of the leptonic component in the wind-MC environment as compared to the homogeneous environment. However, this shortcoming may well simply reflect a limitation of the one-dimensional density structure, especially lacking small-scale features or multi-dimensional effects that must play important roles. 

Our results also provide predictions for future spectral observations.  Our model predictions will be useful to discriminate between different models for Tycho's environments. We estimate that the inhomogeneous environment investigated in the present work will make rapid change in the spectral signatures starting in $\sim$20~yrs, unlike the smooth evolution expected for the homogeneous environment. 


\begin{acknowledgements}
This work is supported by JSPS grant Nos. JP19K03913 (S.H.L.), 24K07092 (S.H.L.), JP19H01936 (T.T.), JP21H04493 (T.T.), JP20H00174 (K.M.), and 24H01810 (K.M.). R.K. acknowledges support by JST, the establishment of university fellowships towards the creation of science technology innovation, Grant Number JPMJFS2123. S.H.L. acknowledges support by the World Premier International Research Center Initiative (WPI), MEXT, Japan. K.M. acknowledges support by The Kyoto University Foundation. 
\end{acknowledgements}


\bibliography{reference}{}
\bibliographystyle{aasjournal}

\end{document}